\documentclass[aps,showpacs,twocolumn]{revtex4} 
\usepackage[dvips]{graphics,graphicx} 
\begin{document}  
\title{"Bloch oscillations" in the Mott-insulator regime}
\author{Andrey R. Kolovsky}
\affiliation{Max-Planck-Institut f\"ur Physik Komplexer Systeme,
                             D-01187 Dresden, Germany}
\affiliation{Kirensky Institute of Physics, 660036 Krasnoyarsk, Russia}
\date{\today}  
\begin{abstract}
We study the dynamical response of cold interacting 
atoms in the Mott insulator phase to a static
force. As shown in the experiment by M.~Greiner et. al.,
Nature \textbf{415}, 39 (2002), this response has resonant
character, with the main resonance defined by
coincidence of Stark energy and on-site interaction 
energy. We analyse the dynamics of the atomic momentum distribution, 
which is the quantity measured in the experiment, for 
near resonant forcing. The momentum distribution
is shown to develop a recurring interference pattern, with
a recurrence time which we define in the paper. 
\end{abstract}  
\pacs{PACS: 32.80.Pj, 03.65.-w, 03.75.Nt, 71.35.Lk}  
\maketitle 


{\bf 1. Introduction.}
Recently much attention has been paid to the properties
of Bose-Einstein condensates of cold atoms loaded into 
optical lattices. In particular, the experimental observation of 
the superfluid (SF) to Mott insulator (MI) transition \cite{Grei02}
has caused a great deal of excitement 
in the field. Note, that besides demonstrating the SF-MI
quantum phase transition, the same experiment also rose
the problem of the system's response to a static
force (used in the experiment to probe the system).
Obviously, the response depends on whether the atoms are initially 
prepared in the SF state or the MI state. The former case was 
investigated theoretically in recent papers \cite{prl,preprint,pre}
(see also related studies \cite{Berg98, Choi99,Chio00}).
It was found that, similar to the case of non-interacting
atoms, the static force induces Bloch oscillations of
the atoms which, however, may be affected rather dramatically by 
the presence of atom-atom interactions. The latter case of MI
initial state was analysed in Ref.~\cite{Brau02,Sach02},
and is also the subject of the present Brief Report.
In particular, we address the evolution of the atomic momentum
distribution not discussed so far. We show that, in
formal analogy with usual Bloch oscillations, 
a static force causes oscillations of the atomic momentum, 
however, with a different characteristic frequency.

\begin{figure}   
\center   
\includegraphics[width=8.5cm, clip]{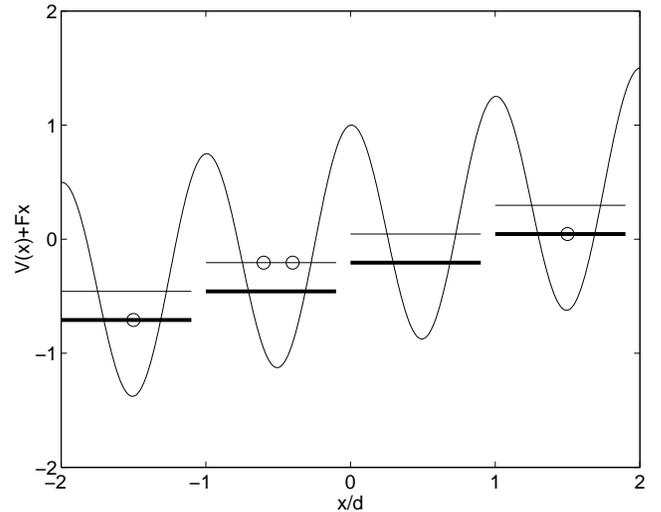}   
\caption{Schematic presentation of the dipole state.
Provided the Stark energy (mismatch between the 
`bold levels' in the figure) is equal to the interaction energy 
(distance between the `bold' and `thick' levels)
the atom may resonantly tunnel in a neighbouring  well,
thus creating particle-hole excitations of the MI state.}    
\label{fig1}   
\end{figure}

\bigskip
{\bf 2. The model and numerical approach.}
Like in our earlier studies \cite{prl,preprint,pre},
we model cold atoms loaded into an optical lattice by 
the Bose-Hubbard Hamiltonian, with an additional Stark term:
\begin{eqnarray}  
\widehat{H} & = & -\frac{J}{2}\left(\sum_l  
\hat{a}^\dag_{l+1}\hat{a}_l  + h.c.\right) \nonumber \\ 
 & &  
+\frac{W}{2}\sum_l\hat{n}_l(\hat{n_l}-1)+dF\sum_l l\hat{n}_l \;. 
\label{1}  
\end{eqnarray} 
In Eq.~(\ref{1}) $J$ is the hopping matrix element, $W$ the on-site
interaction energy, $d$ the lattice period, and $F$ the  magnitude 
of the static force. Throughout the paper we consider a one-dimensional 
lattice and assume, for simplicity, that the filling factor (number of
atoms per lattice site) equals unity. Then, the 
assumption of a Mott-insulator initial state implies
$J<0.17W$. In what follows, however, we shall be mainly interested
in the limiting case $J\ll W$. Under this condition,
the excitation of the system is only possible
if the Stark energy is approximately equal to the interaction energy.
Indeed, for $dF\cong W$ the atoms may resonantly tunnel in the
neighbouring  well, thus forming `dipole' (in the terminology
of Ref.~\cite{Sach02}) states (see Fig.~\ref{fig1}). 

\begin{figure}   
\center   
\includegraphics[width=8.5cm, clip]{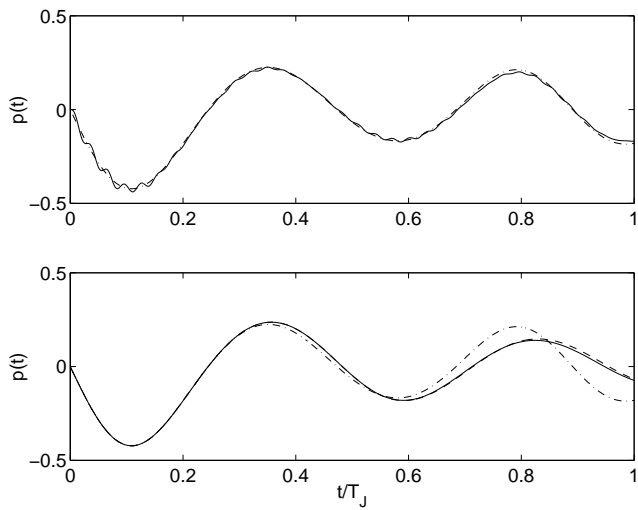}   
\caption{Upper panel: Dynamics of the normalised mean momentum 
($p(t)\rightarrow p(t)/JN$) for $L=N=8$, $W=0.0324$, $dF=W$, and $J=0.0019$.
The solid line shows the exact dynamics, dashed line corresponds to
the resonant approximation. Time axis is scaled with respect to
$T_J=2\pi\hbar/J$, which is the characteristic time-scale of the system. 
Lower panel: Dynamics of the mean momentum calculated 
within the resonant approximation for different system sizes  -- 
$L=8$ (dot-dashed line), $L=12$ (dashed line), and $L=16$ (solid line).
It is seen that, for the considered time interval $t\le T_J$, 
convergence towards the thermodynamic limit is reached already for
$L=12$.}    
\label{fig2}   
\end{figure}

The first step of our analysis is to identify the resonant subspace 
in the system's Hilbert space (spanned by Fock states), i.e. the  
manifold of states resonantly coupled to the MI state. For example, 
for a finite lattice with $L=8$ and $F>0$, the MI state $|11111111\rangle$ 
is coupled to the one-dipole states $|20111111\rangle$, $|12011111\rangle$,
etc., which are coupled to two-dipole states $|20201111\rangle$, 
$|20120111\rangle$, etc., which in turn are coupled to three-dipole states, 
and so on. If $J\ll W$, one actually can neglect the other (non-resonant) 
states when considering the excitation of the system. 
The validity of this resonant approximation is illustrated in 
the upper panel of Fig.~\ref{fig2}, where we compare the time
evolution of the mean atomic momentum, calculated by using the
complete basis (for chosen $L=8$ the total dimension of Hilbert
space is ${\cal N}=6435$), and restricted to the resonant manifold
respectively (${\cal N}_R=47$). It is seen that the resonant approximation
works pretty well already for $J/W\cong 0.05$. It is also worth
noting that, in the resonant approximation and after scaling time
$t\rightarrow Jt/\hbar$, the only relevant parameter of the system 
is the dimensionless detuning 
\begin{equation}
\lambda=(W-dF)/J \;.
\label{detuning}
\end{equation}
%
 
Let us briefly comment on our choice of periodic (cyclic) boundary 
conditions used throughout this paper. These are imposed on (\ref{1}) 
after a gauge transformation which leads to the time-dependent Hamiltonian
\begin{equation}  
\widehat{H}(t)=-\frac{J}{2}\left(e^{i\omega_B t}\sum_l  
\hat{a}^\dag_{l+1}\hat{a}_l  + h.c.\right)
+\frac{W}{2}\sum_l\hat{n}_l(\hat{n_l}-1) \;, 
\label{2}  
\end{equation} 
with $\omega_B=dF/\hbar$ the Bloch frequency. Note that
for periodic boundary conditions the quasimomentum is a conserved
quantity and, hence, the dipole states can be excited only
in coherent superpositions, with the same quasimomentum $\kappa=0$
as for the initial MI state. In particular, for one-dipole 
excitations this constraint defines the state
\begin{equation}
|D(1)\rangle=\frac{1}{\sqrt{L}}\sum_{l=1}^L \widehat{S}^l
|2011\ldots11\rangle \;,
\label{3}
\end{equation}
where $\widehat{S}$ denotes cyclic permutations:
$\widehat{S}|2011\ldots11\rangle=|1201\ldots11\rangle$.
It is worth noting that for two-dipole (three-dipole, etc.)
excitations there are many different states $|D(2)\rangle$,
not related to each other by cyclic permutation. 
In what follows, we refer to the states  $|D(m)\rangle$
as the translationally invariant dipole states.

We conclude this section by a remark on the thermodynamic limit
$L\rightarrow\infty$. Obviously, the dynamics of a system of
finite size differs from the one of an infinite system. However,
this difference emerges only after a finite `correspondence' time.
This is illustrated in the lower panel of Fig.~\ref{fig2}, 
which shows the mean momentum $p(t)$ for different lattice size
$L=8,12,16$. It is seen that an increase of the system size above 
$L=12$ does not change the result and, hence, for the considered
time interval $t\le T_J$ convergence of thermodynamic limit has 
been reached.


\bigskip
{\bf 3.Results of numerical simulations.}
This section reports the results of numerical simulations
the system dynamics obtained within resonant approximation.
In our numerical simulations we followed the scheme of 
present-days laboratory experiments, where one measures the momentum 
distribution of the atoms by using the `free-flight' technique. 
Precisely, after preparation of the initial state (cooling stage),
the atoms are subject to a static force for a given time 
interval $t$ (evolution stage). Then the static field, as
well as the optical potential, is abruptly switched off and
the atoms move in free space. Finally, the spatial distribution 
of the atoms is measured, which carries information about the momentum 
distribution at the end of the evolution stage. Repeating the experiment 
for different time intervals $t$, one recovers the time-evolution
of the momentum distribution $P(p,t)$.
\begin{figure}   
\center   
\includegraphics[width=8.5cm, clip]{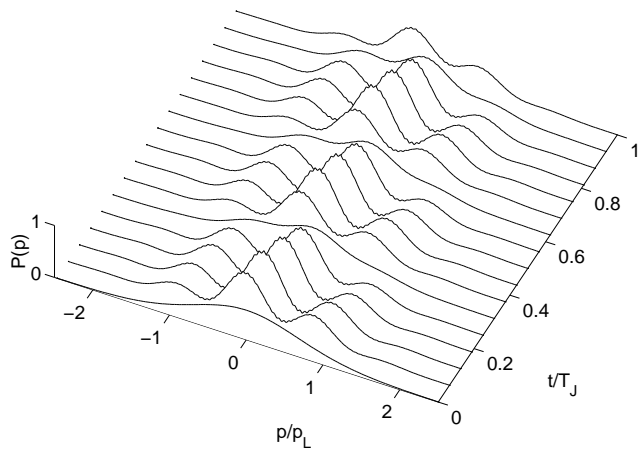}   
\caption{Static force induced dynamics of the atomic momentum 
distribution for cold atoms in a one-dimensional optical lattice. 
The depth of optical the potential is 22 recoil energies, 
the dimensionless detuning $\lambda=2.084$. 
A periodic change  of the distribution is clearly observed. 
Feeble oscillations of the distribution are an artifact, due to 
the finite system size ($L=16$).} 
\label{fig3}   
\end{figure}
 
\begin{figure}   
\center   
\includegraphics[width=8.5cm, clip]{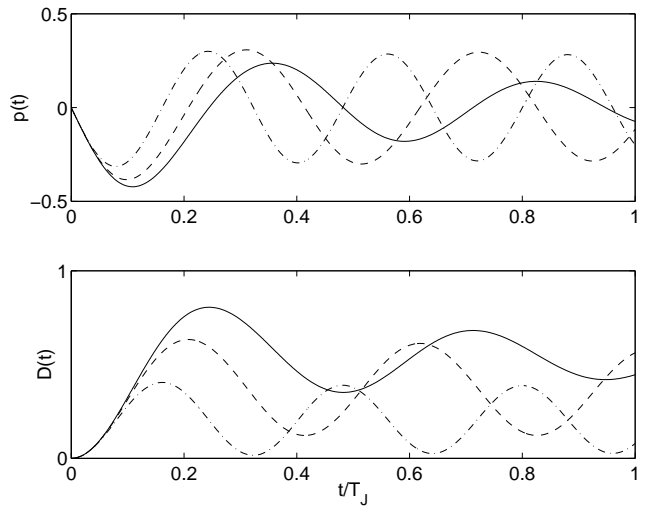}   
\caption{Mean atomic momentum (upper panel) compared to the
average number of dipole states (lower panel), as the functions of time, 
for different values of the detuning $\lambda=0$ (solid line), $1.042$ 
(dashed line), and $2.084$ (dot-dashed line).}
\label{fig4}   
\end{figure}

Figure \ref{fig3} shows the time-evolution of the atomic momentum
distribution for an optical potential depth $v$ equal 
to 22 recoil energies, and a static force strength $F$
corresponding to $\lambda=2.084$. Note that the $v$
uniquely defines the hopping matrix element $J$  ($J=0.0038$ recoil 
energies, for $v=22$) and, hence, the tunnelling time $T_J=2\pi\hbar/J$. 
The amplitude $v$ also defines explicite
form of the Wannier states $\psi_l(x)=\psi_0(x-dl)$ and, thus, 
the initial distribution $P_0(p)=P(p,t=0)$ of the atomic
momenta. Indeed, since the initial state is MI state, 
$|\Psi(t=0)\rangle=|D(0)\rangle\equiv|11\ldots11\rangle$, the momentum 
distribution at $t=0$ is simply the squared Fourier transform 
of the Wannier state, as can be easily derived from the definition 
of the one-particle density matrix \cite{remark1}:
\begin{eqnarray} 
\label{4}
\rho(x,x')&=&\sum_{l,m}\rho_{l,m}(t)\psi_l(x)\psi_m(x') \\
\rho_{l,m}(t)&=&\langle\Psi(t)|\hat{a}^\dag_l\hat{a}_{m}|\Psi(t)\rangle \;.  
\end{eqnarray}
As time evolves, the momentum distribution repeatedly develops
a fringe-like interference pattern. More formally,
\begin{equation} 
\label{5}
P(p,t)=P_0(k) f(p,t) \;,  
\end{equation}
where $f(p,t)=f(p+p_L,t)$ is a periodic function of the momentum with
the period $p_L=2\pi\hbar/d$ defined by the inverse lattice period. For
currently considered case $\lambda=2.084$, the function (\ref{5})
is also (almost) periodic in time with the period $T_\lambda\approx0.33 T_J$. 
Note, however, that $f(p,t)$ is in general not periodic (or quasiperiodic)
in time, although some characteristic time scale prevails.
This statement is illustrated by the upper panel of Fig.~\ref{fig4},
where the temporal  behaviour of the mean momentum is displayed for
different values of the detuning $\lambda$.
One clearly observes the decaying oscillations of the mean momentum,
where both the period of oscillations and the decay rate increase as 
the detuning is decreased. 


\bigskip
{\bf 4.Quasienergy spectrum approach.}
From the point of view of Quantum Optics, momentum  oscillations 
of cold atoms is a process of subsequent excitations of the 
translationally invariant dipole states 
$|D(0)\rangle\leftrightarrow|D(1)\rangle\leftrightarrow\{|D(2)\rangle\}
\leftrightarrow\ldots$.
As an overall characteristic of this process we consider the mean
number of dipole states
\begin{equation} 
\label{6}
D(t)=-\frac{2}{L}\sum_{m=0}^{L/2}\sum_{j} m|c_{m,j}(t)|^2 \;,  
\end{equation}
where the $|c_{m,j}(t)|^2$ are occupation probabilities for different
dipole states. (Note that $D(t)\le1$, due to the chosen normalisation.)
The dynamics of $D(t)$ for three different values of 
the detuning $\lambda$ is shown in the lower panel of Fig.~\ref{fig4}.
A strong correlation between the number of excited dipole states
and oscillations of the mean momentum is clearly observed. 

The above results of our numerical simulations can be qualitatively
understood by analysing the quasienergy spectrum of the system.
An explicite form of the effective Hamiltonian, who's eigenvalues
define the quasienergy spectrum, immediately follows from (\ref{2})
by employing the resonant approximation and is given in Ref.~\cite{Sach02}.
Namely, using the notion of dipole creation operator
$\hat{d}^\dag_l=(1/\sqrt{2}) \hat{a}^\dag_{l-1}\hat{a}_l$ ( which
creates a `hole' at site $l$ and a `quasiparticle' at site $l-1$)
the effective Hamiltonian reads
\begin{equation}
\widehat{H}_{eff}=\lambda\sum_l \hat{d}^\dag_l \hat{d}_l
-\frac{1}{\sqrt{2}}\sum_l\left( \hat{d}^\dag_l+\hat{d}_l\right) \;,
\label{7}
\end{equation}
with the constraint that neither there can be more than one dipole 
at one site ($\hat{d}^\dag_l\hat{d}_l\le1$), nor
two dipoles at neighbouring sites (hard core repulsion).
As noticed in Ref.~\cite{Sach02}, the eigenvalue problem
for the Hamiltonian (\ref{7}) can be mapped to the energy
spectrum problem for a 1D chain of interacting spins, 
where a number of analytical results is known. In particular,
there is an Ising quantum critical point at $\lambda_c=-1.850$,
with qualitatively different ground state of the spin system
below and above this value of $\lambda$. It should be noted,
however, that in the context of our present problem
(dynamical response of the system), the eigenvalue 
problem for the effective Hamiltonian  (\ref{7}) defines the 
{\em quasienergy} spectrum, where the notion of the ground state 
has no physical meaning.
\begin{figure}   
\center   
\includegraphics[width=8.5cm, clip]{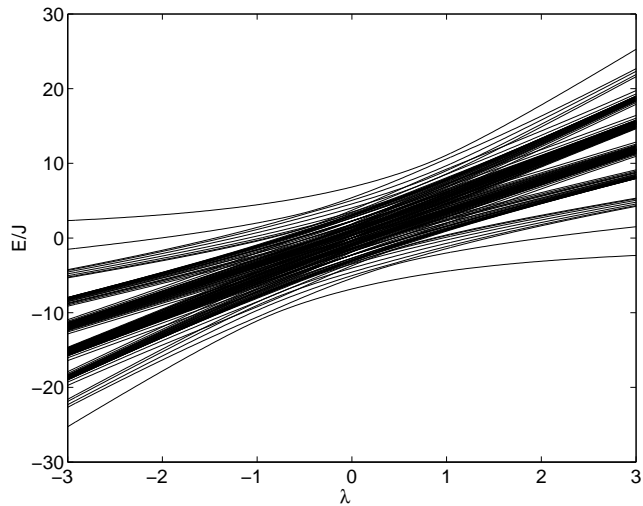}   
\caption{Evolution of the spectrum of effective
Hamiltonian (\ref{7}) under variation of the dimensionless detuning 
$\lambda$. (Lattice size $L=16$.)} 
\label{fig5}   
\end{figure}

To understand the characteristic structure of the quasienergy
spectrum it is convenient to discuss it first for finite $L$.
The result of a direct numerical diagonalisation of $\widehat{H}_{eff}$ 
for $L=16$ is presented in Fig.~\ref{fig5}. This figure shows the 
position of the quasienergy levels, as a function of the 
dimensionless detuning $\lambda$. To avoid possible confusion with 
a similar figure in Ref.~\cite{Sach02}, we note that here only 
the states of the same translational symmetry as the MI state
(i.e., the states with zero value of the quasimomentum) are shown.
It is also worth mentioning that the spectrum has reflection symmetry and,
hence, when discussing the dynamics (rather than thermodynamics), only the
case $\lambda\ge0$ needs to be considered.

Let us discuss the quasienergy spectrum in more detail. It is convenient to 
start with large positive $\lambda$. For a large $\lambda$ the spectrum
consists of separate levels (or bunches of levels), which
in the formal limit $\lambda\rightarrow\infty$ can be associated
with the dipole states (or family of dipole states) $|D(m)\rangle$
with given $m$ ($m\le L/2$). The lowest level in Fig.~\ref{fig5}
is obviously the MI state, the level above the one-dipole state 
$|D(1)\rangle$, followed by the family of states  $|D(2)\rangle$, etc.
(For negative $\lambda$ the situation is reversed and the MI state
$|D(0)\rangle$ is associated with the most upper level.) The key
feature of the spectrum is the finite gap $\Delta$ between the quasienergy 
level $|D(0)\rangle$ and the rest of the spectrum, exiting for arbitrary 
values of $\lambda$. It is precisely this gap, what defines 
the characteristic frequency of atomic oscillations seen in Fig.~\ref{fig4}. 
In the thermodynamic limit we have $\Delta\approx 1.43$ for $\lambda=0$ and 
$\Delta\rightarrow|\lambda|$ for $|\lambda|\rightarrow\infty$ 
\cite{remark2}. Let us also note that in the thermodynamic limit
and for $\lambda\approx0$ the remainder of the quasienergy spectrum
is continuous and gapless. This explains an irreversible decay of
oscillations, although the decay rate (and its $\lambda$-dependence) 
remains an open problem.


\bigskip
{\bf 5.Conclusion.}
We considered the response of the Mott-insulator phase
of cold atoms in an optical lattice to a `resonant' static force.
Here the term `resonant' means that the Stark energy $dF$ approximately
coincides with on-site interaction energy $W$. Under this
condition, the atoms can tunnel in the neighbouring wells of the
optical potential, thus creating particle-hole excitations of
the Mott-insulator state (the `dipole states'). 
This process directly affects the atomic momentum distribution,
which is usually measured in laboratory experiments. 
Namely, the momentum distribution repeatedly develops
an interference pattern with a characteristic period
$T_\lambda$. This period is uniquely defined by the tunnelling
time $T_J=2\pi\hbar/J$ ($J$ is the hopping matrix element) and the
energy gap between two lowest quasienergy levels, which, in turn, is
a unique function of the dimensionless detuning $\lambda=(W-dF)/J$.
In particular, for $\lambda=0$, one has $T_\lambda\approx 0.67 T_J$, 
what, for e.g., for sodium atoms ($E_R/\hbar=2\pi\times8.9$ kHz) 
in optical potential of 22 recoil energies ($J/E_R=0.0038$)
corresponds to approx. 20 milliseconds. 



\begin{thebibliography}{10}
\bibitem{remark1}
Note, by passing, that due to translational invariance of the problem,
the density matrix $\rho_{l,m}(t)$ is a cyclic matrix (i.e. the rows
of the matrix are related to each other by cyclic permutation).
\bibitem{remark2}
Note that the limit of large $|\lambda|$ should be considered with 
precaution because it can violate the resonant approximation.
\end{thebibliography}
\end{document}